# Statistical and Topological Summaries Aid Disease Detection for Segmented Retinal Vascular Images

John T. Nardini[1], Charles W. J. Pugh[2], Helen M. Byrne[2,3]

1. Department of Mathematics and Statistics, The College of New Jersey, Ewing, NJ, 08628, USA
2. Mathematical Institute, University of Oxford, Oxford, Oxfordshire, OX2 6GG, United Kingdom
3. Ludwig Institute for Cancer Research, Nuffield Department of Medicine, University of Oxford, Oxford, Oxfordshire, OX3 7DQ, United Kingdom

**Correspondence**
John T. Nardini, PhD. Department of Mathematics and Statistics, The College of New Jersey, Ewing, NJ, 08628, USA. Email: Nardinij@tcnj.edu

**Abstract**
Disease complications can alter vascular network morphology and disrupt tissue functioning. Diabetic retinopathy, for example, is a complication of types 1 and 2 diabetes mellitus that can cause blindness. Microvascular diseases are assessed by visual inspection of retinal images, but this can be challenging when diseases exhibit silent symptoms or patients cannot attend in-person meetings. We examine the performance of machine learning algorithms in detecting microvascular disease when trained on statistical and topological summaries of segmented retinal vascular images. We apply our methods to three publicly-available datasets and find that, among the 13 total descriptor vectors we consider, either a statistical Box-counting descriptor vector or a topological Flooding descriptor vector achieves the highest accuracy levels on these datasets. We then created a fourth dataset by merging several datasets: the Box-counting vector outperforms all descriptors on this dataset, including the topological Flooding vector which is sensitive to differences in the annotation styles within the combined dataset. Our work represents a first step to establishing which computational methods are most suitable for identifying microvascular disease as well as some of their current limitations. In the longer term, these methods could be incorporated into automated disease assessment tools.

_Keywords:_ microvascular network morphology, segmented image quantification, fractal analysis, topological data analysis, persistent homology, machine learning, disease prediction.

# 1 Introduction

Blood vessel networks deliver nutrients, and remove waste, to maintain tissue homeostasis[1]. Disease complications can alter vascular network morphology, which may lead to insufficient oxygen levels or the accumulation of waste products that disrupt tissue function[2,3]. Many systemic and ocular diseases are associated with altered vessel morphology, including diabetic retinopathy (DR), glaucoma, occlusion, and Coat's disease[4,5]. For example, DR is a common complication of both types 1 and 2 diabetes mellitus (T1DM and T2DM) that stems from high blood sugar levels, which damage small retinal blood vessels[6–8]. DR is the leading cause of blindness in the United States for individuals between the ages of 20 and 64[9,10]. Due to the high prevalence of T1DM and T2DM worldwide, DR incidence is expected to increase over the next 20 years[11]. Current therapeutic options for DR include intravitreal drug injection and laser photocoagulation treatment. Since these treatments can be dangerous and may be ineffective, they are typically only administered to patients with severe DR progression[12].

To prevent DR, the United States and United Kingdom recommend that patients diagnosed with T1DM or T2DM attend annual eye exams at which physicians image patients' retinas using fundus photography. These images are then visually inspected for signs of DR, which include leaking blood vessels, edema, capillary non-perfusion, and damaged nerve tissue[2,13]. In extreme cases, more advanced imaging techniques can be used to provide a more detailed view of the retinal vasculature. While these examinations are effective in preventing DR severity, DR incidence is still projected to increase due to patients exhibiting silent DR symptoms, low exam compliance, and a lack of patient access to health care[11,14]. Thus, there is a growing need for the development of widely-accessible screening tools that can detect early signs of DR as a means to reduce its economic and healthcare impact on the global population. In this study, we will investigate several publicly-available medical image datasets; DR is the most common disease classification in these datasets.

Computational methods that quantify vascular morphology represent a promising tool to aid microvascular disease detection. The development of automated methods that can detect changes in patient microvascular morphology during disease progression is an ongoing area of research. Multiple metrics summarizing static vascular features have been proposed to improve microvascular disease detection, including artery and vessel diameter ratios, vessel tortuosity, and average daughter vessel branching angle[2,13]. The fractal dimension (Df) descriptor is also widely used to quantify network complexity for microvascular disease detection. Fractals are self-similar spatial patterns that repeat across multiple spatial scales. Popovic et al. showed that Df values increase with DR severity on the Standard Diabetic Retinopathy Database Calibration Level 1 dataset[15,16]. The effectiveness of the Df descriptor for DR detection is limited, however, as in this study, the Df descriptor was only applied to images of retinal vascular networks with many branches. Further, there is no consensus about how Df varies with microvascular disease severity: several studies have reported that Df increases with increasing DR severity[17,18] while others report that it decreases with increasing DR severity[19–21]. We note that the Df descriptor has been widely used to describe morphological changes in a range of biological systems[22,23], and that care is needed to ensure that there is sufficient spatial resolution in the data to calculate the Df. Indeed, some studies have suggested that the Df may have been inappropriately used in some neuroscience and ecological applications[24,25]. In practice, the Df is computed from a vector which we term the Box-counting vector; here, we consider both the scalar-valued Df and the Box-counting vector for our analyses.

Topological data analysis (TDA) provides an alternative computational method to aid disease detection. TDA is a recent field of mathematics that uses concepts from topology and geometry to infer the structure of data[26,27]. Persistent homology (PH) is a commonly-used area of TDA that identifies topological features (e.g., connected components and loops) that are characteristic of a dataset[28]. In contrast to scalar-valued summary statistics, the output from PH summarizes features in the data across multiple scales. Additionally, previous theoretical results guarantee that PH computations are robust to noisy and/or incomplete data[29]. TDA summaries have been used to detect disease from medical data, with applications in melanoma[30], breast cancer[31], pulmonary disease[32], heart disease[33], and Covid-19[34,35]. Several of these studies combine TDA summaries with machine learning algorithms to cluster, or classify, the data into groups that share similar topological characteristics and disease status[35,36]. Disease prediction is then performed by computing the same topological summaries on unseen data and identifying the cluster to which each data point is most likely to belong.

Thus, TDA and, in particular, PH computations represent a promising multiscale approach to describe vascular network morphology, which may aid microvascular disease detection. PH has been applied to vascular network data previously: Bendich et al. showed that PH summaries can reveal the effects of aging on brain vascular morphology[37], and Stolz et al. used PH to uncover effects of radiotherapy on tumor vascular structure that were not seen using statistical methods[38]. We recently applied PH computations and a machine learning algorithm to cluster a large collection of simulated vessel networks according to their topological features[39]. We found that simulations within the same clusters were generated from similar input model parameters. We propose that both statistical and PH computations of segmented retinal vascular network images, can be combined with machine learning algorithms to detect disease in medical images. While we focus on segmented vascular images, Garside et al. combined PH computations of colored fundus images with machine learning to detect DR[39,40]. Their prediction algorithms required many input summaries to achieve high accuracy levels. We apply our methods to detect the presence of general microvascular disease on four publicly-available datasets and are able to achieve similar accuracy levels with only one statistical or topological summary vector.

In this work, we show that both statistical and topological computations provide summaries that can aid general microvascular disease detection from segmented vascular images. DR is the most common disease classification in these data. We apply these methods to four existing and publicly-available datasets, namely the STructured Analysis of the REtina (STARE)[41], Digital Retinal Images for Vessel Extraction (DRIVE)[42], Child Heart And health Study in England (CHASE)[43], and High Resolution Fundus (HRF)[44] datasets. Each dataset provides two-dimensional binary image segmentations that were manually annotated by experts from fundus images and a disease classification for each image. We train support vector machines to predict disease classification from either statistical or topological computations and use 5-fold cross validation to estimate the accuracy of each approach. We find that both statistical and topological descriptor vectors can achieve high accuracy levels on these datasets: a topological Flooding descriptor achieves the highest accuracy level on the STARE datasets, and the statistical Box-counting vector performs best on the HRF dataset. When we combine data from all four datasets into an "All" dataset, the same statistical descriptor vector leads to the highest accuracy levels. Further analysis of the Flooding descriptor vector reveals that it is sensitive to how each vessel segmentation is annotated. This work represents a first step to establishing which computational methods are most suitable for identifying microvascular disease. In the longer term, the statistical and topological descriptor vectors could be integrated to develop automated tools that aid disease diagnosis and assist clinical decision making, while also making healthcare more effective, accessible, and affordable.

## 2 Materials and Methods

We obtained two-dimensional binary vessel segmentation images (VSIs) and their disease classifications from four publicly available datasets. We summarize the morphology of each VSI by computing standard and topological descriptor vectors. These vectors are then used to train a machine learning algorithm that predicts VSI disease classification. By comparing the algorithm's accuracy when trained with different types of input vectors, we identify those descriptor vectors that are most informative for disease prediction. The study pipeline is summarized in Figure 1 and consists of three steps:

> Step 1. Available data,
> Step 2. Data analysis, and
> Step 3. Disease prediction.

All code used in our implementation uses publicly-available python software and is available at https://github.com/johnnardini/VSI_automated_disease_detection.git.

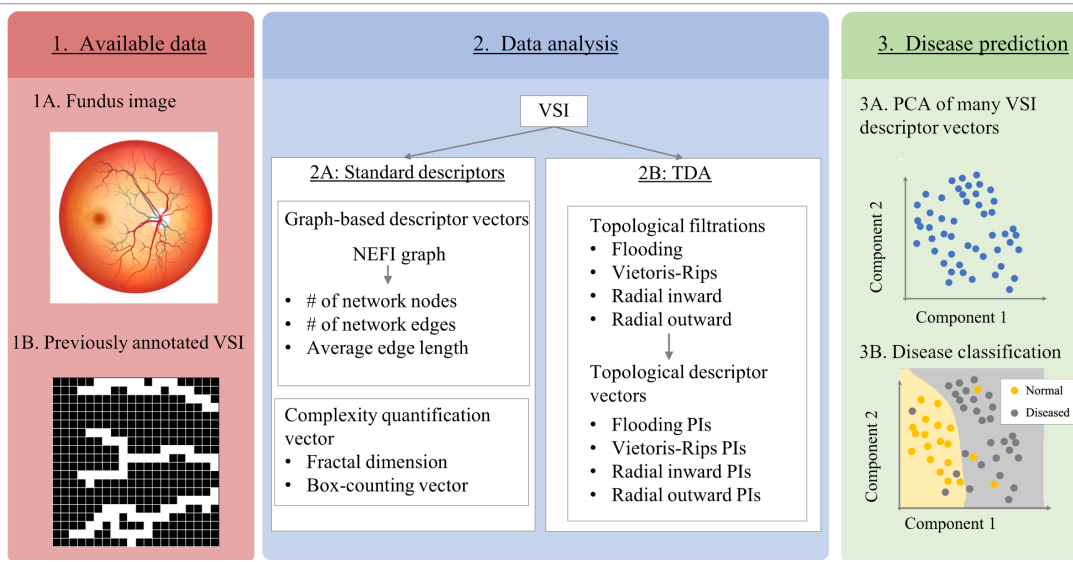

Figure 1: Overview of methods pipeline. Step 1. Available data. Two-dimensional fundus images, their previously annotated binary vessel segmentation images (VSIs), and disease classifications were extracted from four publicly-available datasets. Step 2. Data analysis. We compute standard descriptor vectors and topological descriptor vectors for each VSI. A) Standard descriptors. We represent each VSI as a graph using the Network Extraction From Images (NEFI) software. We count the number of nodes and edges in the graph, and compute its average edge length. We also compute the fractal dimension for each VSI. B) Topological data analysis (TDA). We construct Flooding, Vietoris-Rips, Radial inward, and Radial outward filtrations for each VSI. Each filtration yields two two-dimensional persistence images (PIs) that describe the connected components and loops of the filtration. The PIs are then vectorized. Step 3. Disease prediction. VSI disease classification is predicted using either a standard descriptor vector or a vectorized PI from Step 2. A) Principal components analysis (PCA) is used to project the descriptor vectors onto their two dominant principal components. B) Support vector machines are trained to predict the disease classification of each VSI using either dimension-reduced descriptor vectors or scalar-valued descriptor vectors.

## 2.1 Available data

VSIs and their disease classifications were obtained from the STARE, DRIVE, CHASE, and HRF datasets. Each VSI has previously been manually annotated from fundus images by expert inspection. VSIs are two-dimensional binary images, in which pixel values of one denote the presence of blood vessels and pixel values of zero denote their absence. The technical aspects of each dataset are summarized in Table 1.

Table 1: Summary of datasets used throughout the study.

|  | STARE | DRIVE | CHASE | HRF | All |
|---|---|---|---|---|---|
| Number of patients | 20 | 20 | 14 | 45 | n/a |
| VSIs per patient | 2 | 1 | 4 | 1 | n/a |
| Total number of VSIs | 40 | 20 | 56 | 45 | 161 |
| Disease classes | Normal (n = 20), Diseased (n = 20) | Normal (n=17), Diseased (n=3) | n/a | Normal (n=15), Diseased (n = 30) | Normal (n = 108), Diseased (n =53) |

STARE, STructured Analysis of the REtina; DRIVE, Digital Retinal Images for Vessel Extraction; CHASE, Child Heart and Health Study in England; HRF, High Resolution Fundus; DR, Diabetic Retinopathy; VSI, Vessel segmentation image.

**2.1.1 The STARE dataset** This dataset contains fundus images from 20 patients[41]. Each image was annotated separately by two experts, and has size 700x605 pixels. Of the 20 patients, 10 were diagnosed as 'Normal', and 10 were diagnosed with a specific disease, such as Hollenhorst emboli and vein occlusion. For simplicity, we classify the latter patients as 'Diseased'. The STARE data are available at https://cecas.clemson.edu/ ahoover/stare/.

**2.1.2 The DRIVE dataset** This dataset includes fundus images from 20 patients[42]. Each image was annotated once and has size 565x584 pixels. Of the 20 patients, 17 were diagnosed as Normal and 3 as Diseased. The DRIVE data are available at https://drive.grand-challenge.org/.

**2.1.3 The CHASE dataset** This dataset contains fundus images from the left and right retinas of 14 school children[43]. Each image was annotated separately by two experts, and has size 999x960 pixels. No diagnoses were made, so we classify all children as Normal. The CHASE data are available at https://blogs.kingston.ac.uk/retinal/chasedb1/.

**2.1.4 The HRF dataset** This dataset includes fundus images from 45 patients[44]. Each image was annotated once and has size 3,304x2,336 pixels. Of the patients, 15 were diagnosed as Normal, and 30 were diagnosed as Diseased. The HRF data are available at https://www5.cs.fau.de/research/data/fundus-images/.

**2.1.5 The All dataset** We do not perform disease prediction on the DRIVE or CHASE datasets since they contain few Diseased images. Instead, we combine the VSIs from all four datasets into an "All" dataset that contains 161 VSIs, with 109 diagnosed as Normal and 52 as Diseased. To address inconsistencies between datasets, each VSI in the "All" dataset has been cropped about its field of view and resized to a common size of 700x605 pixels. The STARE and DRIVE datasets do not provide fields of view for each image; we estimated each VSI's field of view from these datasets by determining the smallest rectangle that contains all nonzero pixels.

**2.2 Data analysis**

**2.2.1 Standard descriptors** We summarize VSIs by computing three graph-based and two complexity vectors and refer to them collectively as *standard descriptor vectors*.
In order to compute the graph-based standard descriptor vectors, we first use the Network Extraction From Images (NEFI) software to generate a graph for each VSI[45]. As shown in Figure 2, we place nodes at locations where three vessel segments intersect and also at terminal points of vessel segments. Edges connect pairs of nodes that are connected by a vessel segment in the VSI. We use the resulting graph to compute three network quantities: the number of nodes, the number of edges, and the average length of each edge.

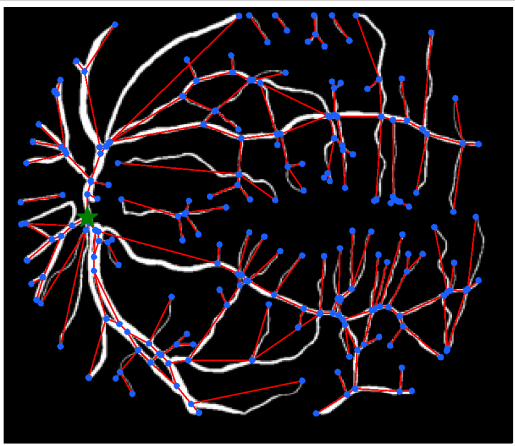

Figure 2: An illustrative graph representation of a vessel segmentation image (VSI). A graph generated with the Network extraction from images (NEFI) software superimposed over the input VSI from the STARE dataset. Graph nodes and edges are depicted by blue dots and red lines respectively, and the central node is indicated by the green star. For this VSI, the computed standard descriptor values are: # nodes = 224, # edges = 234, Avg. edge length = 38.38 pixel lengths, and fractal dimension = 1.64.

To create the two complexity-based standard descriptor vectors, we compute the fractal dimension (Df) of each VSI using the Box-counting method[15]. An illustration of the Box-counting method is shown in Supplementary Figure 1. This method decomposes the VSI into boxes of pre-determined side lengths. We consider ten box side lengths, $s$: for the smallest value of $s$, each box is the size of one pixel; for the largest value, two boxes cover the shorter side of the VSI. For each value of $s$, boxes are assigned a value of one if they contain at least one nonzero pixel; otherwise they are assigned a value of zero. We denote by $N(s)$ the number of boxes of size $s$ with value one. We assume that $ln(N(s))$ is linearly proportional to $ln(s)$, in which case the slope of the line of best fit between $s$ and $ln(N(s))$ approximates the Df of a given VSI[46]. The two complexity quantities we consider are the Df and the Box-counting vector, $ln(N(s))$.

**2.2.2 Topological data analysis** TDA is an emerging field of research that combines concepts from algebraic topology and computational geometry to analyze the shape of high dimensional data. PH is a widely-used methodology within TDA[26–28]. In contrast to descriptors that summarize data at a single scale, PH quantifies features of the data over many scales by varying a specific scale parameter (for VSIs, suitable scale parameters include distance from a vessel segment or distance from the optic disk).

PH achieves a multiscale description of data by constructing *filtrations*. From a range of values of the scale parameter, a filtration generates a sequence of embedded graph-like structures on the data points. These structures are sparsely connected for small values of the scale parameter and highly connected for large values. PH quantifies topological features in the filtration. The topological features we consider are connected components and loops (loops form when connections between data points enclose an empty region). Features that are present across a large range of scale parameter values are called *persistent* and may be characteristic of a dataset. The outputs from PH are barcodes and persistence diagrams, which visualize the existence and persistence of topological features in the data. In order to compare PH outputs, perform statistical analyses, and/or apply classification methods from machine learning, persistence diagrams can be transformed into two-dimensional persistence images which can, in turn, be converted into vectors (i.e. vectorized)[47].

We investigate how disease alters retinal morphology by computing four filtrations for each VSI: the Vietoris-Rips, Flooding, Radial outward, and Radial inward filtrations. Before describing each filtration, we present an illustrative example that highlights the key concepts of PH.

**Persistent Homology: an example** We illustrate our PH pipeline (Figure 4) by applying the Flooding filtration to the VSI from step 1B of Figure 1.

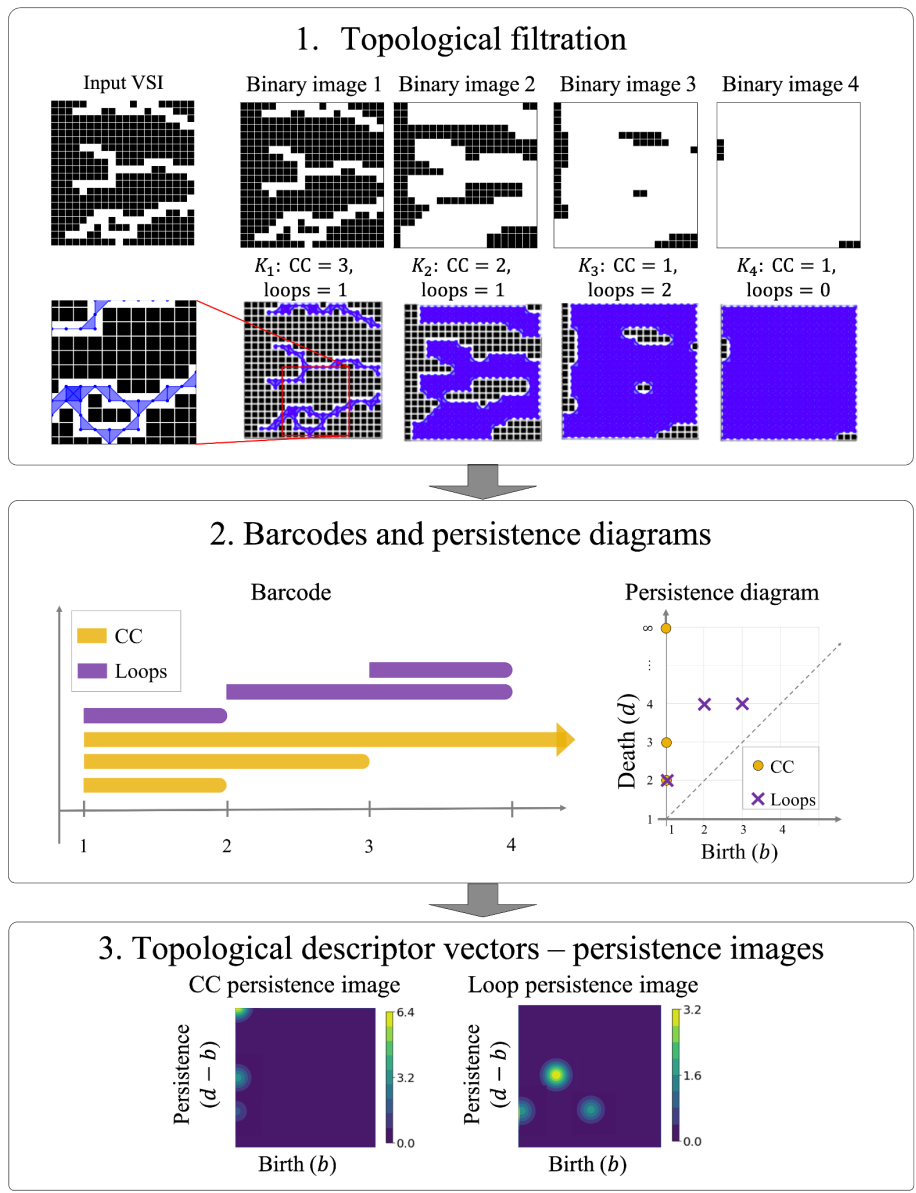

Figure 3: Persistent homology (PH) pipeline to quantify vessel segmentation image (VSI) morphology. The Flooding filtration is applied to the VSI from step 1B of Figure 1.
1) Topological filtration. A sequence of binary images, starting with the initial VSI, is constructed by applying a Flooding step to the previous image. Each binary image is then converted into a simplicial complex, $K_i$, for $i = 1,2,3,4$. PH quantifies the lifetimes of connected components (CC) and loops within the filtration $K = \{K_1, K_2, K_3, K_4\}$. 2) Barcodes and persistence diagrams. The lifetime of each topological feature type is summarized by a barcode or persistence diagram. 3) Topological descriptor vectors - persistence images. Persistence diagrams are converted into two-dimensional persistence images, a form of topological descriptor vector that is suitable for machine learning tasks.

In part 1 of the PH pipeline, we compute the Flooding filtration by constructing a sequence of two-dimensional binary images. The first binary image identifies the vessel segments present in the VSI as white pixels, the second identifies the original vessel segments and their nearest neighbors as white pixels, while the third identifies the original segments and their first and second nearest neighbors as white pixels, and so on. The process terminates when all pixels are white (for the VSI in Figure 4, this corresponds to binary image 5, not shown). We apply four Flooding steps in this example, but typically more steps are needed to analyze VSIs from retinal images.

The sequence of binary images is then converted into a filtration by replacing each binary image with a set of points, lines, and triangles, known collectively as a *simplicial complex*. The $i^{th}$ ($i = 1, 2, 3, 4$) simplicial complex, $K_i$, is generated from the $i^{th}$ binary image in the following way: a point is placed at the centroid of each white pixel, lines are drawn between any two neighboring points, and filled in triangles are included where three neighboring points form a right triangle (see inset of $K_1$ in Figure 4). The filtration is given by $K = \{K_1, K_2, K_3, K_4\}$.

We use PH to quantify the lifetime, or persistence, of topological features (connected components and loops) in $K$. A feature is born in filtration step $b$ if it first appears in $K_b$; the feature dies on filtration step $d > b$ if it is no longer present in $K_i$, for $i > d$ and its persistence is defined to be $d - b$. For example, in $K_1$, there are three connected components and one loop; in $K_2$, one connected component dies (because two connected components from $K_1$ join together), one loop dies (because it becomes filled in), and a new loop is born; in $K_3$, one connected component dies, and a new loop is born; finally, in $K_4$, a single connected component persists, and both previous loops die. Any additional Flooding steps result in an entirely white binary image, whose simplicial complexes contain one connected component and no loops.

In part 2 of the PH pipeline, we summarize the lifetimes of the topological features using barcodes and persistence diagrams. Within a *barcode*, each feature is represented by a bar of length $d - b$ placed in the interval $[b, d)$. For the example in Figure 4, the barcode of connected components is given by $\{[1,2), [1,3), [1,\infty)\}$, and the barcode of loops is given by $\{[1,2), [2,4), [3,4)\}$. *Persistence diagrams* are an alternative way to depict the topological features; each persistence interval $[b, d)$ is represented by a two-dimensional point with co-ordinates $(b, d)$.

In part 3, we convert persistence diagrams into two-dimensional *persistence images* (PIs). The birth-death co-ordinates, $(b, d)$, of topological features are replaced by a two-dimensional Gaussian distribution, centered at $(b, d - b)$ with constant variance ($\sigma = 1.0$). PIs are created by multiplying each Gaussian distribution by its persistence, $d - b$, and then summing over all weighted distributions. The resulting function is then converted into a discretized image, with a fixed grid size that we chose to be $N_{PI} \times N_{PI} = 50 \times 50$. PIs for connected components and loops are created separately and can be converted into vectors to facilitate subsequent analyses and comparison of VSIs.

**Topological filtrations** There is currently no consensus about which topological filtrations are best suited for determining how disease alters retinal morphology and, in turn, predicting the presence of disease from VSI data. To address this issue, we consider four filtrations: the Flooding filtration, the Vietoris-Rips filtration, and the "Radial outward" and "Radial inward" filtrations (Figure 4).

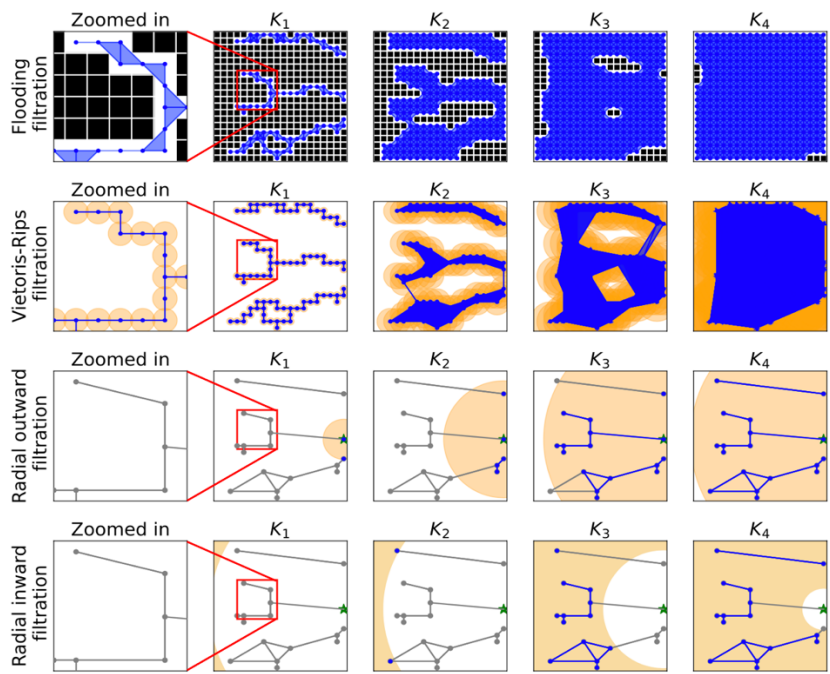

Figure 4: Overview of the four topological filtrations that are used to analyze VSI data. First row: the Flooding filtration creates a sequence of simplicial complexes from a sequence of binary images. Second row: the Vietoris-Rips filtration creates a sequence of simplicial complexes from a point cloud by connecting points that are within a given radius of each other. Third row: the Radial outward filtration creates a sequence of simplicial complexes from a graph by including nodes within a given radius of the graph's central node (denoted with a green star). Fourth row: the Radial inward filtration creates a sequence of simplicial complexes from a graph by including nodes outside of a given radius of the graph's central node.

For each filtration, we define N values of its scale parameter; from the $i^{th}$ ($i = 1, ..., N$) value, we build a simplicial complex, $K_i$. A simplicial complex is a structure consisting of points, lines, triangles, and their higher order counterparts that connect data points together. We refer to the resulting sequence of simplicial complexes, $K = \{K_1, K_2, K_3, K_4\}$, as a filtration.

The Flooding filtration: We generate a sequence of two-dimensional binary images, starting with the original VSI (Figure 4, top row). We create a new binary image by looping over all nonzero pixels in the previous binary image. In addition to holding fixed the values of all nonzero pixels, the values of their eight neighboring pixels are set to one. This process is repeated until a sequence of 70 binary images has been generated, which fills each VSI. We retain the odd-numbered images (resulting in a final sequence of 35 images) to ease subsequent computation. To construct the Flooding filtration, $K^{flood}$, we convert the $i^{th}$ binary image into the simplicial complex, $K_i^{flood}$ ($i = 1, 2, ..., 35$), by placing points at the centroids of all nonzero pixels, drawing lines between any two points that are within the eight surrounding pixels of each other, and placing filled-in triangles where three connected points form a right triangle.

The Vietoris-Rips filtration: We convert each VSI into a two-dimensional point cloud by placing a point at the centroid of each nonzero pixel (Figure 4, second row). Since the Vietoris-Rips filtration is computationally expensive, we subsample the point cloud to reduce computational time: we use a Greedy furthest point sampling algorithm to extract 2000 representative points from the original point cloud data (i.e., approximately 10% of the points)[48]. We compute the Vietoris-Rips filtration (and the corresponding persistence images for connected components and loops) for 50 separate subsamplings and then save the averaged persistence images for connected components and loops.

To construct the Vietoris-Rips filtration, $K^{VR}$, for a given point cloud, we define $N = 40$ distance values $0 \leq \epsilon_1 < \epsilon_2 < \cdots < \epsilon_N$ and place a circle of radius $\epsilon_i$ around each subsampled point. We connect, with a line, any two points whose circles overlap, and connect with a filled-in triangle any three points whose circles pairwise overlap. Each $K_i^{VR}$ ($i = 1, 2, \ldots, N$) is generated by combining all such points, lines, and filled-in triangles that result from $\epsilon_i$. We choose $\epsilon_i = (i - 1)185/39$ to analyze each dataset.

The Radial outward and Radial inward filtrations: We consider a Radial outward filtration, $K^{outward}$, and a Radial inward filtration, $K^{inward}$ (the Radial outward and inward filtrations are depicted in Figure 4 in the third and fourth rows, respectively). In both cases, for a given VSI, we first compute a graph that comprises a set of nodes, $V$, and a set of edges, $E$ (Figure 2). Nodes are placed at locations where three vessel segments intersect and at terminal points of vessel segments. Edges connect pairs of nodes that are connected by a vessel segment in the VSI. We use the NEFI software to generate each graph[45]. We designate the node from $V$ with the highest betweenness centrality measure as the graph's central node, $v^c$ (see green star in Figure 2). The betweenness centrality measure for each node $v$ is the proportion of shortest paths between all pairs of nodes that pass through $v$. This definition for $v^c$ is chosen to identify a node located near the optic disk of each VSI.

To construct $K^{outward} = \{K_1^{outward}, \ldots, K_{40}^{outward}\}$, we consider $N = 40$ radial values $0 \leq r_1 < r_2 < \cdots < r_{40}$. All nodes from $V$ located within distance $r_i$ of $v^c$ are included as points in $K_i^{outward}$. If two nodes included in $K_i^{outward}$ are connected by an edge from $E$, then we connect them with a line in $K_i^{outward}$. We do not include filled-in triangles in $K_i^{outward}$. For the HRF dataset; we choose $r_i = (i - 1)3000/39$ pixel lengths; for all other datasets, we choose $r_i = (i - 1)700/39$ pixel lengths. These values ensure that all computed NEFI graphs are covered by the end of the filtration.

To construct $K^{inward} = \{K_1^{inward}, \ldots, K_{40}^{inward}\}$, we use the same radial values $r_i$ ($i = 1, 2, \ldots, N$) as those used for the Radial outward filtration, and proceed as follows. All nodes from $V$ located at a distance $r_N - r_i$, or greater, from $v^c$ are included as points in $K_i^{inward}$. If two nodes included in $K_i^{inward}$ are connected by an edge in $E$, then we connect them with a line in $K_i^{inward}$. We do not include filled in triangles in $K_i^{inward}$.

**Barcodes and persistence diagrams** PH uses barcodes and persistence diagrams to quantify the lifetimes of topological features, here connected components and loops, in a filtration[26–28, 49, 50]. As discussed in the example, a barcode is a collection of intervals $\{[b_1, d_1), ..., [b_k, d_k)\}$ designating the birth and death times of each topological feature, and a persistence diagram is a collection of two-dimensional points $\{(b_1, d_1), ..., (b_k, d_k)\}$ summarizing the lifetime of each topological feature (Figure 3). We create two separate persistence diagrams from each filtration, one for connected components and one for loops. We use the software Ripser (version 0.4.1) to compute the Vietoris-Rips filtration because it efficiently computes this filtration. Ripser cannot compute the remaining filtrations, so we use Gudhi (version 3.4.1) to compute the Flooding and radial filtrations. Both Ripser and Gudhi are publicly available.

**Persistence images** We require vectors of the same length to represent each VSI for disease prediction. Therefore, we map all persistence diagrams into two-dimensional persistence images (PIs) of the same size and convert the PIs into vectors of a fixed length (Figure 4). The resulting topological descriptor vectors are stable to perturbations in data and can be used for machine learning and data science algorithms [27,39,47].

Briefly, a PI is computed from a persistence diagram in the following way:

1. Each birth-death pair $(b, d)$ is transformed into a birth-persistence pair $(b, p) = (b, d - b)$;
2. Each $(b, p)$ pair is recast as a two-dimensional Gaussian normal distribution, with mean $(b, p)$ and variance $\sigma$ (here, we fix $\sigma = 1.0$);
3. Each Gaussian distribution is discretized to produce a two-dimensional image of size $N_{PI} \times N_{PI}$ (here, we fix $N_{PI} = 50$);
4. All Gaussian images are multiplied by their persistence, $p$, and then summed to produce the PI;
5. The PI is vectorized to produce a topological descriptor vector of length $N_{PI}^2$ that is subsequently used for disease prediction.

More information on PIs and their computation is available in [47]. We compute separate PIs for connected components and loops from each filtration, producing 8 topological descriptor vectors for each VSI.

**2.2.3 Disease prediction** We assess the performance of each standard descriptor and topological descriptor vector in predicting VSI disease classification. Our disease prediction pipeline uses principal components analysis (PCA) to reduce the dimension of the descriptor vectors and support vector machines (SVMs) for disease prediction. Our implementation is available on the Github repository and employs publicly-available software (Sci-kit learn, version 0.24.2).

**Dimensionality reduction** Each Box-counting vector is of length 10 because we consider 10 box side lengths, and each vectorized PI is of length 2,500 because we fix $N_{PI} = 50$. The information contained in these large vectors can often be condensed into smaller vectors. As one example, the Df descriptor assumes that the information content of the Box-counting vector, $ln(N(s))$, can be summarized by the slope of the best-fit line between $s$ and $ln(N(s))$. The PCA algorithm optimizes vector information content by projecting each vector onto a smaller number of directional co-ordinates (or principal components)[51]. The first principal component is the direction that maximizes the variance in the data. Each successive principal component maximizes the remaining variance in the data while being orthogonal to all previously-chosen principal components.

We use PCA to project each descriptor vector onto the 2-dimensional subspace spanned by the first two principal components (see Step 3A of Figure 1). PCA is performed on all topological descriptor vectors and the Box-counting vector; PCA is not performed on the remaining standard descriptors because they are scalar-valued. We chose to reduce each vector to a length of 2 because reducing each vector to a length of 3 did not significantly change our subsequent disease classification results (results not shown).

**Disease classification** For each descriptor vector, we use a supervised learning algorithm to predict VSI disease classification. In more detail, we implement SVMs, with linear kernels[52], to identify the curves or surfaces that best partition the data into distinct groups according to disease type (see Step 3B in Figure 1).

In the schematic from Step 3B of Figure 1, we have plotted several two-dimensional vectors as two-dimensional points and colored each point according to its disease classification. An SVM is the curve $y = g(x)$ that best partitions a training portion of the data by disease classification. Details for training SVMs to determine this curve from these data and their classifications are provided in [53]. After training, an SVM can be used to predict the classification of any "new" data points by determining the relationship between each point and the curve $y = g(x)$. For example, in Step 3B of Figure 1, $y = g(x)$ is the curve separating the gray and yellow regions. The SVM will predict that any new point located to the left of the curve $y = g(x)$ has a Normal classification (as marked in yellow) and any other point has a Diseased classification (as marked in gray).

Cross validation is commonly used to assess the accuracy with which machine learning classifiers predict the classification of data not used during model training. We perform 5-fold cross validation to quantify and then compare the accuracy of SVMs associated with each of the standard and topological descriptor vectors. Given $M$ data-classification points $(x_1, y_1), (x_2, y_2), \ldots, (x_M, y_M)$, where $x_i$ is a descriptor vector and $y_i$ is the corresponding disease classification ($y_i = 1$ for Normal, $y_i = 0$ for Diseased), cross validation is performed as follows. During a given round of cross validation, 80% of the data-classification points are placed into a training dataset and the remaining 20% are placed into a testing dataset. An SVM is trained on the training dataset to predict $y_i$ from $x_i$; afterwards, the trained SVM is then used to predict the classification of each $x_i$ from the testing dataset. The SVM's out-of-sample (OOS) accuracy is the percentage of these predictions that match the true classification labels. In 5-fold cross validation, the mean OOS accuracy of the SVM is determined from five different training-testing splits, and each data-classification pair is placed in the testing dataset only once. Since the OOS accuracy score depends on how the data are split, we perform 100 rounds of 5-fold cross validation and report average values and their standard deviations.

# 3 Results

## 3.1 The Box-counting vector is the most accurate standard descriptor of disease

We predicted VSI disease classifications using SVMs associated with each standard descriptor vector for the STARE, HRF and All datasets (Table 2). We treated the two STARE experts' annotated VSI collections as separate datasets. SVMs associated with the Box-counting vector usually outperformed the other standard descriptors, achieving the highest OOS accuracy for all datasets except the STARE expert 2 dataset. Plotting the computed Box-counting vectors for the HRF dataset shows that, for small values of $s$, Normal VSIs have higher computed $N(s)$ values when compared to Diseased VSIs (Figure 5).

Table 2: Summary of the mean out-of-sample (OOS) accuracy scores (standard deviations in parentheses) obtained when standard descriptor vectors are used to train support vector machines (SVMs) to predict the disease status of VSIs from all four datasets. We highlight in bold the highest OOS accuracy for a given dataset. Rows are ordered by vectors with the highest mean OOS accuracy score across all four datasets. Principal components analysis (PCA) was used to reduce each Box-counting vector $ln\ (N(s)\ )$ to two-dimensions before prediction; PCA was not used on the remaining descriptors because they are scalar-valued.

| Standard Descriptors | STARE expert 1 | STARE expert 2 | HRF | All |
|---|---|---|---|---|
| Box-counting | **83 (3.5) %** | 82 (3.0) % | **94 (1.5) %** | **81 (1.0) %** |
| Df | 75 (6.5) % | **84 (4.1) %** | 88 (2.0) % | 65 (1.7) % |
| # Edges | 68 (5.2) % | 75 (6.9) % | 68 (4.4) % | 71 (1.3) % |
| # Nodes | 66 (5.2) % | 71 (6.9) % | 69 (3.3) % | 71 (1.3) % |
| Avg. Edge Length | 54 (7.8) % | 64 (6.0) % | 68 (3.7) % | 67 (2.3) % |

Df, fractal dimension; Avg., average.

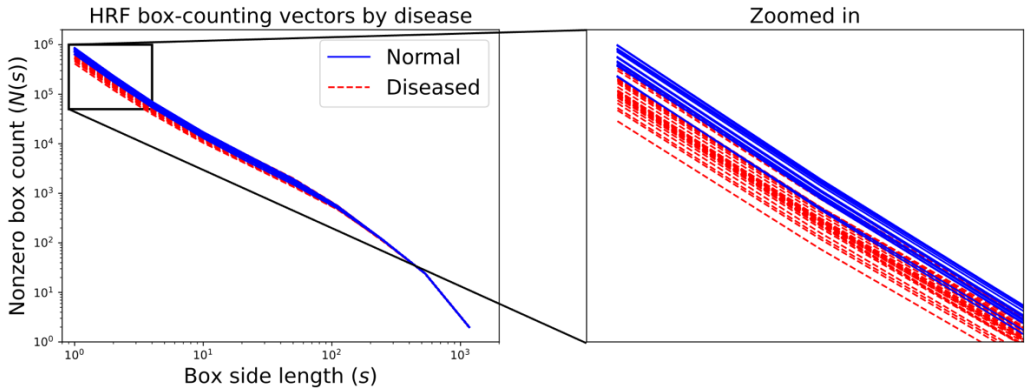

Figure 5: The Box-counting vectors from the HRF dataset. Plots of the computed Box-counting descriptor vectors, $ln(N(s))$, for Normal and Diseased vessel segmentation images from the HRF dataset.

The mean Df values decrease in the presence of disease for the STARE and HRF datasets (Table 3). On the STARE and HRF datasets, the mean Df values for normal and diseased VSIs vary, but the relative percentage reductions (3.2%, 6.1%, 3.0%, respectively) are similar in magnitude. We conclude that, for the diseased retinas contained within the datasets, the associated vascular networks are less complex than those for healthy retinas. On the All dataset, the mean Df value does not change in the presence of disease. The discrepancy between these results may be an artifact of the data: over half of the Diseased images in the All datasets are from the HRF dataset (which has more complex VSIs) while over half of the Normal images are from the STARE, DRIVE, and CHASE datasets (which have simpler VSIs).

Table 3: Average fractal dimension (Df) values for Normal and Diseased VSIs from all four datasets.

|  | STARE expert 1 | STARE expert 2 | HRF | All |
| --- | --- | --- | --- | --- |
| Mean Normal Df value (SD) | 1.55 (0.017) | 1.65 (0.016) | 1.69 (0.016) | 1.61 (0.051) |
| Mean Diseased Df value (SD) | 1.50 (0.041) | 1.55 (0.066) | 1.64 (0.020) | 1.61 (0.043) |

SD, standard deviation.

## 3.2 Topological descriptor vectors accurately (but inconsistently) predict disease classification

We used SVMs associated with each topological descriptor vector to predict VSI disease classifications for all four datasets (Table 4). Forest plots of these summaries are also available (Supplementary Figure 2). On the STARE datasets, SVMs associated with Flooding (loops) PIs perform best and outperform all standard descriptor vectors. However, the Flooding (loops) descriptor does not perform well on the remaining datasets. The SVM trained on the Radial outward (loops) PIs performs best on the HRF dataset, but it does not outperform the Df and Box-counting vectors on this dataset. On the All dataset, the SVM associated with the Radial inward (loops) PIs performs best, but SVMs trained on the Box-counting vector achieve higher accuracy scores on this dataset.

Table 4: Summary of the mean out-of-sample (OOS) accuracy scores (standard deviations in parentheses) obtained when topological descriptor vectors are used to train SVMs to predict the disease status of VSIs from all four datasets. Principal components analysis (PCA) was used to reduce each topological descriptor vector to two-dimensions before prediction. Bold highlights the highest OOS accuracy for a given dataset. Rows are ordered by vectors with the highest mean OOS accuracy score on all four datasets.

| Topological descriptor vectors | STARE expert 1 | STARE expert 2 | HRF | All |
|---|---|---|---|---|
| Flooding (loops) | **90 (1.9) %** | **94 (3.4) %** | 60 (4.7) % | 65 (2.0) % |
| Radial outward (loops) | 84 (2.9) % | 58 (7.8) % | **80 (4.0) %** | 70 (1.6) % |
| Vietoris-Rips (CC) | 78 (3.7) % | 80 (3.5) % | 65 (2.9) % | 65 (1.8) % |
| Vietoris-Rips (loops) | 80 (5.6) % | 80 (3.2) % | 62 (4.6) % | 65 (1.9) % |
| Radial outward (CC) | 62 (7.6) % | 45 (10.7) % | 61 (3.9) % | 68 (1.4) % |
| Radial inward (CC) | 36 (10.0) % | 58 (9.7) % | 63 (4.2) % | 71 (1.8) % |
| Radial inward (loops) | 38 (9.2) % | 47 (6.9) % | 64 (4.2) % | **72 (1.3) %** |
| Flooding (CC) | 59 (8.0) % | 42 (9.0) % | 59 (4.1) % | 70 (1.4) % |

CC, connected components.

## 3.3 Interpretation of the topological descriptor vectors

Projecting the dimension-reduced Flooding (loops) PIs from the STARE expert 1 dataset onto their first two principal components clearly partitions the Normal and Diseased data (Figure 6). Visual inspection of the VSI data reveals structural differences between the two groups: Normal VSIs exhibit highly-branched and intertwined vessel networks, whereas Diseased VSIs contain fewer secondary branches and form loosely-connected vessel networks. The Flooding (loops) PIs are able to summarize these dominant (topological) features. A representative PI from a Normal patient has a high PI density along the vertical line $b = 0$, indicating that many loops are present in the VSI. A representative PI from a Diseased VSI has smaller densities along this line because it has fewer loops. We observe similar results when the dimension-reduced Flooding (loops) PIs from the STARE expert 2 dataset are projected onto their first two principal components (Supplementary Figure 3).

SVMs trained on the dimension-reduced Radial outward (loops) PIs achieve the highest mean OOS accuracy scores on the HRF dataset. Projecting these PIs onto their first two principal components partitions the Normal and Diseased Data (Supplementary Figure 2). In these data, Normal VSIs exhibit more loops than Diseased VSIs. One consequence of this morphological difference is that the Normal PIs have a peak density that occurs near $b = 5$ whereas Diseased PIs have a peak that occurs near $b = 10$. The locations of these peaks suggest that, for Normal VSIs, more loops are located closer to the optic disk than for Diseased VSIs.

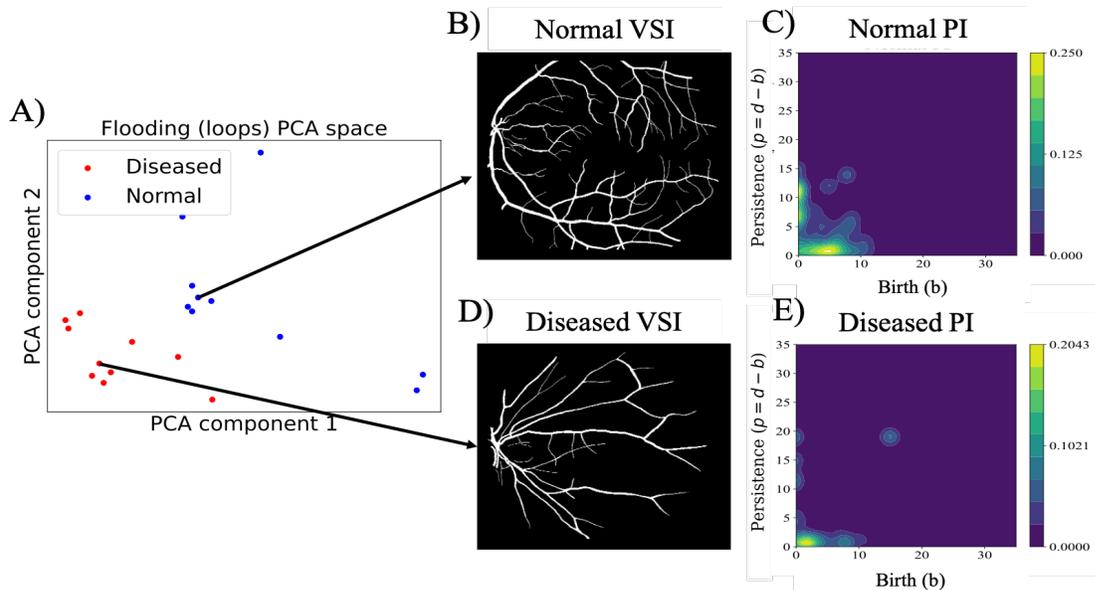

Figure 6: Flooding (loops) PIs partition the STARE expert 1 dataset. A) Plotting the dimension-reduced Flooding (loops) PIs along the first two principal components. B) Representative Normal VSI, and C) Flooding (loops) PI from the representative Normal VSI. D) Representative Diseased VSI, and E) Flooding (loops) PI from the representative Diseased VSI.

**3.4 Topological descriptor vectors are sensitive to VSI annotation styles**

Even though the same fundus images were annotated to create the VSIs from the two STARE datasets, visual inspection of VSIs reveals different morphological patterns are present in the data (Figure 7A-B). For example, expert 1 does not include as many small vessel segments in the VSIs as expert 2. As a result, fewer loops are present in the VSIs from the first expert. VSIs from the HRF dataset also include more vessel segments and loops when compared to VSIs from the STARE expert 1 dataset (Figure 7C).

We investigated the discrepancies in VSI loop counts by computing the number of loops in each VSI for all three datasets by disease classification (Figures 7D-E) and by computing the mean number of loops (Table 5). The mean computed numbers of loops per VSI in the STARE expert 2 and HRF datasets are much higher than the mean values for the STARE expert 1 dataset for both Normal and Diseased data. The presence of disease decreases the mean number of loops in both STARE datasets, however, the mean number of loops in a Diseased VSI from the STARE expert 2 is higher than the mean number of loops in a Normal VSI from the STARE expert 1 dataset. In contrast with the two STARE datasets, the presence of disease leads to an increase in the mean number of loops within the HRF dataset.

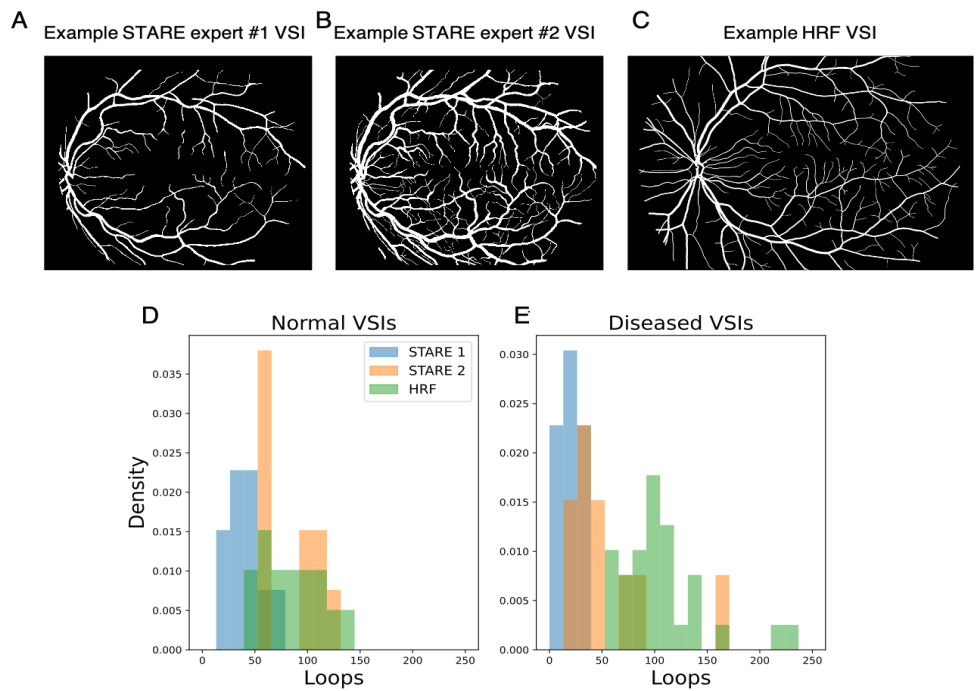

Figure 7: Vessel segmentation images (VSIs) from each dataset contain different numbers of loops. A-C) Example VSIs from the STARE Expert 1, STARE Expert 2, and HRF datasets. The VSIs from the STARE datasets were annotated from the same fundus image. D-E) Histograms summarizing the number of loops present in each VSI from the three datasets for Normal and Diseased data.

Table 5: Mean VSI number of loops by dataset and disease class. Summaries of the mean number of loops in the STARE expert 1, STARE expert 2, and HRF datasets (standard deviations in parentheses).

|  | STARE expert 1 | STARE expert 1 | HRF |
| --- | --- | --- | --- |
| Normal VSIs | 43 (14.7) | 86 (25.8) | 83.2 (29.3) |
| Diseased VSIs | 20 (9.4) | 53 (40.4) | 106.9 (39.8) |

VSI, vessel segmentation image.

## 4 Discussion

We studied the accuracy of machine learning classifiers in predicting general disease status from retinal VSIs when trained with either standard descriptor vectors or topological descriptor vectors. The classifiers achieved the highest predictive accuracy on the HRF and All datasets when trained with either the Box-counting vector or the fractal dimension. On the STARE dataset, the Flooding (loops) descriptor achieved the best results. Further analysis showed that, for three of the four datasets of interest, disease reduces VSI complexity, as quantified by the fractal dimension. We also found that diseased VSIs have fewer loops, as quantified by the Flooding (loops) filtration. These findings point to morphological differences between Normal and Diseased VSIs in the datasets.

In this study, we have shown how statistical and topological methods can aid assessment of general microvascular disease. Each of the top-performing methods have limitations, however, which suggests that more research is needed to determine which metrics (or combinations of metrics) should be used to analyze and classify VSIs. For example, the Box-counting vector performs the best on the HRF and All datasets, but is outperformed by the Flooding (loops) descriptor on both STARE datasets. The Flooding (loops) descriptor does not perform well on the HRF and All datasets. Further analysis of the computed Flooding (loops) descriptors from the All dataset suggests that this method is sensitive to how each VSI is annotated. The fractal dimension and Radial outward (loops) descriptors also perform well, though the accuracy of both descriptors change by at least nine percentage points between the two STARE datasets. One possible explanation for these discrepancies is the small size of the datasets. In future work, we aim to analyze larger datasets, including longitudinal data, to determine how descriptors change with disease progression and to identify methods that perform consistently and robustly across datasets. We aim also to determine how to combine statistical and topological methods to leverage their respective advantages and, in turn, to improve the overall accuracy of disease assessment.

We combined statistical and topological methods with a machine learning algorithm to detect microvascular disease from VSIs. In separate work, Garside et al. applied another TDA method to fundus images in order to detect diabetic retinopathy[40]. Taken together, these studies provide preliminary evidence supporting the potential for combining methods from statistics and TDA with machine learning algorithms to automate detection of retinal diseases from medical images. A strength of TDA is that it is underpinned by mathematical theory, including stability results which guarantee that small changes to input data (e.g., the removal or addition of vessel segments) do not significantly alter the resulting output[29]. However, we found in this work that many topological descriptor vectors were not practically stable, as their performance on the two STARE datasets was not consistent. This finding was surprising because both datasets were annotated from the same fundus images. In future work, we plan to implement extended persistence and multiparameter persistence to determine whether these methods provide more robust quantification of VSIs[54,55].

We note that artificial neural networks are an alternative approach to automate the detection of general microvascular disease[56]. Because deep neural networks perform exceptionally well on computer vision tasks, several recent studies have proposed their use to extract VSIs from fundus images and then to use them for disease classification[57–59]. A concern about the use of neural networks, however, is that it can be challenging to interpret the outputs from these black-box models[60]. Topological and statistical analyses, on the other hand, generate metrics that are interpretable. In future work, we will combine topological and statistical methods with machine learning algorithms and mathematical modeling to infer the mechanisms that lead to diabetic retinopathy and other microvascular diseases of the retina[39,61].

A limitation of this study is the small sample size. We focused on the STARE, DRIVE, CHASE, and HRF datasets due to their wide use in the scientific community and public availability. In the future, we aim to produce a fully automated pipeline that uses neural networks to extract VSIs from fundus images and then uses our methodology for their analysis and classification[62]. This will enable us to analyze larger retinal image datasets and to distinguish different disease types (e.g., Normal, Diabetic retinopathy, or Glaucoma).

In this work, we focused on two-dimensional VSIs, but TDA methods have been used to characterize three-dimensional vascular networks in the brain and in cancer[37,38,63]. These methods could be adapted to analyze three-dimensional, optical coherence tomography angiography images[63].

## 5 Perspective

Our work shows how statistical and topological methods can be combined with machine learning algorithms to predict whether retinal vascular networks are Normal or Diseased. A statistical approach, namely the Box-counting descriptor vector, provides the most accurate classification results over the four datasets considered. A topological descriptor, generated from a Flooding filtration, performed well on some datasets, though we found that this approach is sensitive to how vascular images are annotated. Taken together, these results highlight the importance of testing a range of alternative metrics when designing automated tools to analyze, quantify, and classify biomedical imaging data.

The methods developed in this work could also contribute to the growing use of telemedicine in making healthcare more accessible, affordable and effective. Digital images of a patient's eyes, taken using smartphones, could be sent to clinics for automated analysis and disease classification, prior to review by a medical expert. In addition to delivering remote clinical services to patients, automated diagnoses prior to medical consultations, would enable doctors to devote more time to evaluating non-standard cases.

**Acknowledgements:** The authors thank BJ Stolz, H Adams, L Ziegelmeier, and CM Topaz for helpful discussions and commentary.

**Funding information:** HMB is grateful for the support provided by the UK Centre for Topological Data Analysis EPSRC grant EP/R018472/1. The authors acknowledge use of the ELSA high performance computing cluster at The College of New Jersey for conducting the research reported in this paper. This cluster is funded in part by the National Science Foundation under grant numbers OAC-1826915 and OAC-1828163.

**Conflict of interest**: The authors declare no conflicts of interest.

**Data availability:** All data used in this work is publicly available. All code used in this work is publicly available at https://github.com/johnnardini/VSI_automated_disease_detection.git

# References

1. Davis, G. E., Norden, P. R. & Bowers, S. L. K. Molecular control of capillary morphogenesis and maturation by recognition and remodeling of the extracellular matrix: functional roles of endothelial cells and pericytes in health and disease. *Connect. Tissue Res.* **56**, 392–402 (2015).

2. Kee, A. R., Wong, T. Y. & Li, L.-J. Retinal vascular imaging technology to monitor disease severity and complications in type 1 diabetes mellitus: A systematic review. *Microcirculation* **24**, (2017).

3. MacCormick, I. J. C. *et al.* Spatial statistical modelling of capillary non-perfusion in the retina. *Sci. Rep.* **7**, 16792 (2017).

4. Sen, M., Shields, C. L., Honavar, S. G. & Shields, J. A. Coats disease: an overview of classification, management and outcomes. *Indian J. Ophthalmol.* **67**, 763–771 (2019).

5. Morris, B., Foot, B. & Mulvihill, A. A population-based study of Coats disease in the United Kingdom I: epidemiology and clinical features at diagnosis. **24**, 1797–1801 (2010).

6. Falavarjani, K. G. *et al.* Spatial distribution of diabetic capillary non-perfusion. *Microcirculation* **28**, e12719 (2021).

7. Wong, T. Y. *et al.* Retinal microvascular abnormalities and their relationship with hypertension, cardiovascular disease, and mortality. *Surv. Ophthalmol.* **46**, 59–80 (2001).

8. Wong, T. Y. *et al.* Quantitative retinal venular caliber and risk of cardiovascular disease in older persons: the cardiovascular health study. *Arch. Intern. Med.* **166**, 2388–2394 (2006).

9. Engelgau, M. M. *et al.* The evolving diabetes burden in the United States. *Annals of Internal Medicine* vol. 140 945 (2004).

10. Tapp, R. J. *et al.* The prevalence of and factors associated with diabetic retinopathy in the Australian population. *Diabetes Care* **26**, 1731–1737 (2003).

11. Kalogeropoulos, D., Kalogeropoulos, C., Stefaniotou, M. & Neofytou, M. The role of tele-ophthalmology in diabetic retinopathy screening. *J. Optom.* **13**, 262–268 (2020).

12. Stitt, A. W. *et al.* The progress in understanding and treatment of diabetic retinopathy. *Prog. Retin. Eye Res.* **51**, 156–186 (2016).

13. Wu, L., Fernandez-Loaiza, P., Sauma, J., Hernandez-Bogantes, E. & Masis, M. Classification of diabetic retinopathy and diabetic macular edema. *World J. Diabetes* **4**, 290–294 (2013).

14. Ogurtsova, K. *et al.* IDF diabetes atlas: global estimates for the prevalence of diabetes for 2015 and 2040. *Diabetes Res. Clin. Pract.* **128**, 40–50 (2017).

15. Popovic, N. *et al.* Fractal characterization of retinal microvascular network morphology during diabetic retinopathy progression. *Microcirculation* e12531 (2019).

16. Kauppi, T. *et al.* The DIARETDB1 diabetic retinopathy database and evaluation protocol. *Procedings of the British Machine Vision Conference 2007* (2007) doi:10.5244/c.21.15.

17. Cheung, N. *et al.* Quantitative assessment of early diabetic retinopathy using fractal analysis. *Diabetes Care* **32**, 106–110 (2009).

18. Lim, S. W. *et al.* Retinal vascular fractal dimension and risk of early diabetic retinopathy: A prospective study of children and adolescents with type 1 diabetes. *Diabetes Care* **32**, 2081–2083 (2009).

19. Broe, R. *et al.* Retinal vessel calibers predict long-term microvascular complications in type 1 diabetes: the Danish Cohort of Pediatric Diabetes 1987 (DCPD1987). *Diabetes* **63**, 3906–3914 (2014).

20. Broe, R. *et al.* Retinal vascular fractals predict long-term microvascular complications in type 1 diabetes mellitus: the Danish Cohort of Pediatric Diabetes 1987 (DCPD1987). *Diabetologia* **57**, 2215–2221 (2014).

21. Broe, R. Early risk stratification in pediatric type 1 diabetes. *Acta Ophthalmol.* **93 Thesis 1**, 1–19 (2015).

22. Lorthois, S. & Cassot, F. Fractal analysis of vascular networks: insights from morphogenesis. *J. Theor. Biol.* **262**, 614–633 (2010).

23. Gazit, Y., Berk, D. A., Leunig, M., Baxter, L. T. & Jain, R. K. Scale-invariant behavior and vascular network formation in normal and tumor tissue. *Phys. Rev. Lett.* **75**, 2428–2431 (1995).

24. Murray, J. D. Use and abuse of fractal theory in neuroscience. *The Journal of Comparative Neurology* vol. 361 369–371 (1995).

25. Halley, J. M. *et al.* Uses and abuses of fractal methodology in ecology. *Ecology Letters* vol. 7 254–271 (2004).

26. Carlsson, G. Topology and data. *Bulletin of the American Mathematical Society* vol. 46

255–308 (2009).

27. Otter, N., Porter, M. A., Tillmann, U., Grindrod, P. & Harrington, H. A. A roadmap for the computation of persistent homology. *EPJ Data Sci* **6**, 17 (2017).

28. Ghrist, R. Barcodes: The persistent topology of data. *Bulletin of the American Mathematical Society* vol. 45 61–76 (2007).

29. Cohen-Steiner, D., Edelsbrunner, H. & Harer, J. Stability of persistence diagrams. *Discrete & Computational Geometry* vol. 37 103–120 (2007).

30. Koseki, K. *et al.* Assessment of skin barrier function using skin images with topological data analysis. *NPJ Syst Biol Appl* **6**, 40 (2020).

31. Nicolau, M., Levine, A. J. & Carlsson, G. Topology based data analysis identifies a subgroup of breast cancers with a unique mutational profile and excellent survival. *Proc. Natl. Acad. Sci. U. S. A.* **108**, 7265–7270 (2011).

32. Belchi, F. *et al.* Lung topology characteristics in patients with chronic obstructive pulmonary disease. *Scientific reports* **8**, 5341 (2018).

33. Aljanobi, F. A. & Lee, J. Topological data analysis for classification of heart disease data. *2021 IEEE International Conference on Big Data and Smart Computing* (2021) doi:10.1109/bigcomp51126.2021.00047.

34. Hajij, M., Zamzmi, G. & Batayneh, F. TDA-Net: fusion of persistent homology and deep learning features for COVID-19 detection from chest x-ray images. *Conference proceedings - IEEE engineering in medicine and biology society* **2021**, 4115–4119 (2021).

35. Hickok, A., Needell, D. & Porter, M. A. Analysis of spatial and spatiotemporal anomalies using persistent homology: case studies with COVID-19 data. *arXiv [physics.soc-ph]* (2021).

36. Singh, Mémoli & Carlsson. Topological methods for the analysis of high dimensional data sets and 3d object recognition. *Eurographics Symposium on Point-Based Graphics* (2007) doi:10.2312/spbg.spbg07.091-100/091-100.

37. Bendich, P., Marron, J. S., Miller, E., Pieloch, A. & Skwerer, S. Persistent homology analysis of brain artery trees. *The Annals of Applied Statistics* vol. 10 (2016).

38. Stolz, B. J. *et al.* Multiscale topology characterizes dynamic tumor vascular networks. *Science Advances* vol. 8 (2022).

39. Nardini, J. T., Stolz, B. J., Flores, K. B., Harrington, H. A. & Byrne, H. M. Topological data analysis distinguishes parameter regimes in the Anderson-Chaplain model of angiogenesis. *PLOS Computational Biology* vol. 17 e1009094 (2021).

40. Garside, K., Henderson, R., Makarenko, I. & Masoller, C. Topological data analysis of high resolution diabetic retinopathy images. *PLoS One* **14**, e0217413 (2019).

41. Hoover, A. D., Kouznetsova, V. & Goldbaum, M. Locating blood vessels in retinal images by piecewise threshold probing of a matched filter response. *IEEE Transactions on Medical Imaging* vol. 19 203–210 (2000).

42. Staal, J., Abràmoff, M. D., Niemeijer, M., Viergever, M. A. & van Ginneken, B. Ridge-based vessel segmentation in color images of the retina. *IEEE Trans. Med. Imaging* **23**, 501–509 (2004).

43. Fraz, M. M. *et al.* An ensemble classification-based approach applied to retinal blood vessel segmentation. *IEEE Trans. Biomed. Eng.* **59**, 2538–2548 (2012).

44. Budai, A., Bock, R., Maier, A., Hornegger, J. & Michelson, G. Robust vessel segmentation in fundus images. *International Journal of Biomedical Imaging* vol. 2013 1–11 (2013).

45. Dirnberger, M., Kehl, T. & Neumann, A. NEFI: network extraction from images. *Sci. Rep.*

**5**, 15669 (2015).

46. Schleicher, D. Hausdorff dimension, its properties, and its surprises. *Am. Math. Mon.* **114**, 509–528 (2007).

47. Adams, H. *et al.* Persistence images: a stable vector representation of persistent homology. *Journal of Machine Learning Research* **18**, 1–35 (2017).

48. Bhaskara, Vadgama & Xu. Greedy sampling for approximate clustering in the presence of outliers. *Adv. Neural Inf. Process. Syst.* (2019).

49. Edelsbrunner, H. & Harer, J. L. *Computational topology: an introduction*. (American Mathematical Society, 2022).

50. H Edelsbrunner, J. L. H. Persistent homology: a survey. *Contemp. Math.* **453**, 257–282 (2008).

51. Abdi, H. & Williams, L. J. Principal component analysis. *Wiley Interdiscip. Rev. Comput. Stat.* **2**, 433–459 (2010).

52. Boser, B. E., Guyon, I. M. & Vapnik, V. N. A training algorithm for optimal margin classifiers. *Proceedings of the fifth annual workshop on Computational learning theory - COLT '92* (1992) doi:10.1145/130385.130401.

53. Scholkopf, B. *et al.* Comparing support vector machines with Gaussian kernels to radial basis function classifiers. *IEEE Trans. Signal Process.* **45**, 2758–2765 (1997).

54. Thorne, T., Kirk, P. D. W. & Harrington, H. A. Topological approximate Bayesian computation for parameter inference of an angiogenesis model. *Bioinformatics* (2022) doi:10.1093/bioinformatics/btac118.

55. Vipond, O. *et al.* Multiparameter persistent homology landscapes identify immune cell spatial patterns in tumors. *Proc. Natl. Acad. Sci. U. S. A.* **118**, (2021).

56. Wang, F., Casalino, L. P. & Khullar, D. Deep learning in medicine—promise, progress, and challenges. *JAMA Internal Medicine* vol. 179 293 (2019).

57. Li, Y.-H., Yeh, N.-N., Chen, S.-J. & Chung, Y.-C. Computer-assisted diagnosis for diabetic retinopathy based on fundus images using deep convolutional neural network. *Mobile Information Systems* vol. 2019 1–14 (2019).

58. Abdelsalam, M. M. Effective blood vessels reconstruction methodology for early detection and classification of diabetic retinopathy using OCTA images by artificial neural network. *Informatics in Medicine Unlocked* vol. 20 100390 (2020).

59. Zhao, H., Sun, Y. & Li, H. Retinal vascular junction detection and classification via deep neural networks. *Comput. Methods Programs Biomed.* **183**, 105096 (2020).

60. Ching, T. *et al.* Opportunities and obstacles for deep learning in biology and medicine. *Journal of the Royal Society, Interface* **15**, (2018).

61. Fu, X., Gens, J. S., Glazier, J. A., Burns, S. A. & Gast, T. J. Progression of diabetic capillary occlusion: a model. *PLoS Comput. Biol.* **12**, e1004932 (2016).

62. Alom, M. Z., Hasan, M., Yakopcic, C., Taha, T. M. & Asari, V. K. Recurrent residual convolutional neural network based on U-Net (R2U-Net) for medical image segmentation. *arXiv [cs.CV]* (2018).

63. Li, M. *et al.* Image projection network: 3D to 2D image segmentation in OCTA images. *IEEE Trans. Med. Imaging* **39**, 3343–3354 (2020).

# Supplementary Information: Statistical and Topological Summaries Aid Disease Detection for Segmented Retinal Vascular Images

John T. Nardini[1], Charles W. J. Pugh[2], Helen M. Byrne[2,3]

1. Department of Mathematics and Statistics, The College of New Jersey, Ewing, NJ, 08628, USA
2. Mathematical Institute, University of Oxford, Oxford, Oxfordshire, OX2 6GG, United Kingdom
3. Ludwig Institute for Cancer Research, Nuffield Department of Medicine, University of Oxford, Oxford, Oxfordshire, OX3 7DQ, United Kingdom

**Correspondence**
John T. Nardini, PhD. Department of Mathematics and Statistics, The College of New Jersey, Ewing, NJ, 08628, USA. Email: Nardinij@tcnj.edu

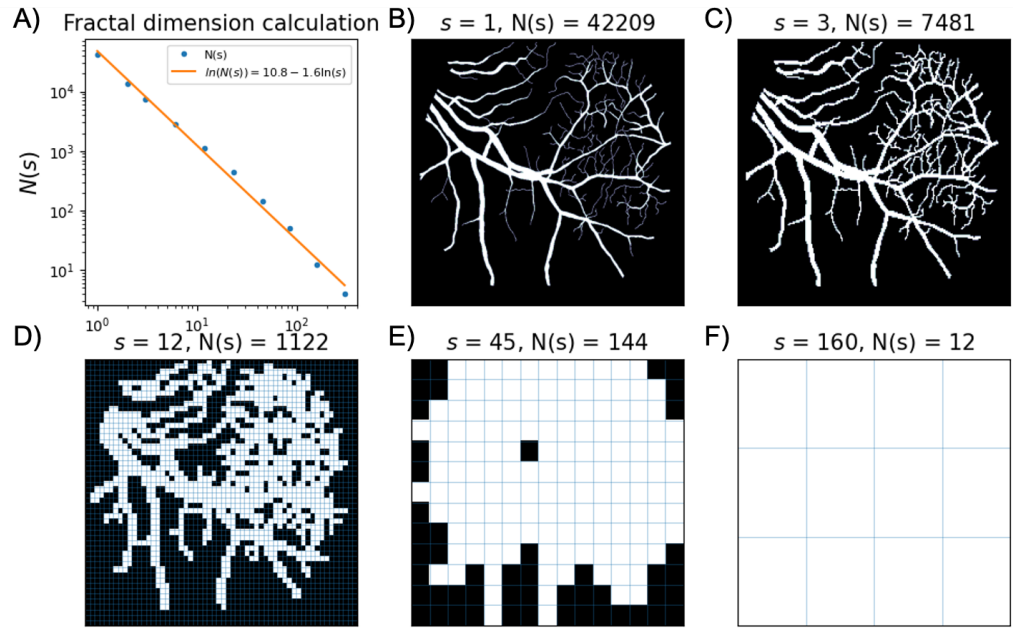

Supplementary Figure 1: Computation of the fractal dimension (Df) for a vessel segmentation image (VSI). A) For several box side length values, $s$, the VSI is split into boxes of size $s \times s$, and $N(s)$ denotes the number of boxes with at least one nonzero pixel inside of them. A line is fit to $\ln \ln (N(s))$ against $\ln \ln (s)$. The Df of the VSI is computed by multiplying the slope of this best-fit line by -1. B-F) The boxes with at least one nonzero pixel inside of them are shown in white for box side lengths of 1, 3, 12, 45, and 160. The blue lines are the boundaries of each box; these lines are not depicted in panels B & C for visual clarity.

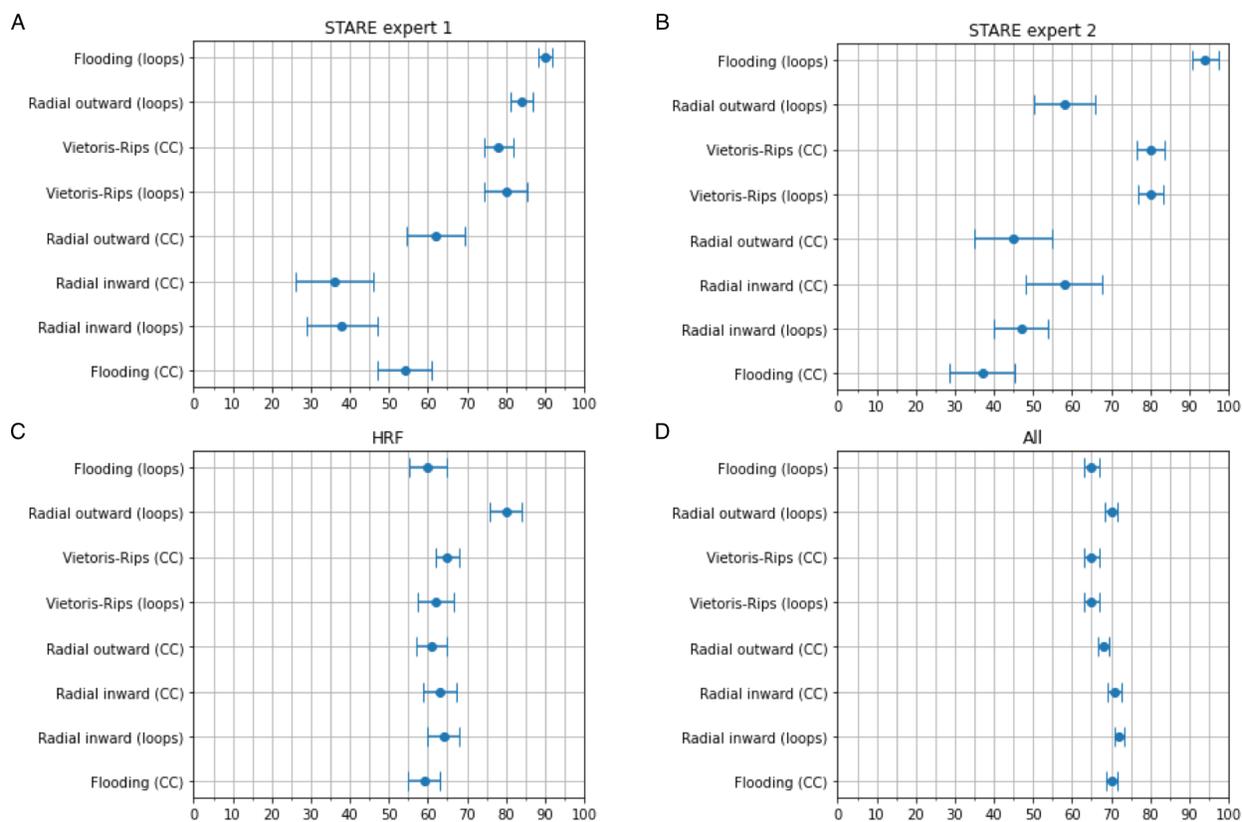

Supplementary Figure 2: Forest plot summary of the mean out-of-sample (OOS) accuracy scores obtained when topological descriptor vectors are used to train SVMs to predict the disease status of VSIs from all four datasets. Plots are made for the A) STARE expert 1, B) STARE expert 2, C) HRF, and D) All datasets. The dot in each row denotes the mean OOS accuracy score obtained over 100 rounds of 5-fold cross validation. The upper and lower bounds in each row denote the mean value plus or minus one standard deviation, respectively.

CC, connected components.

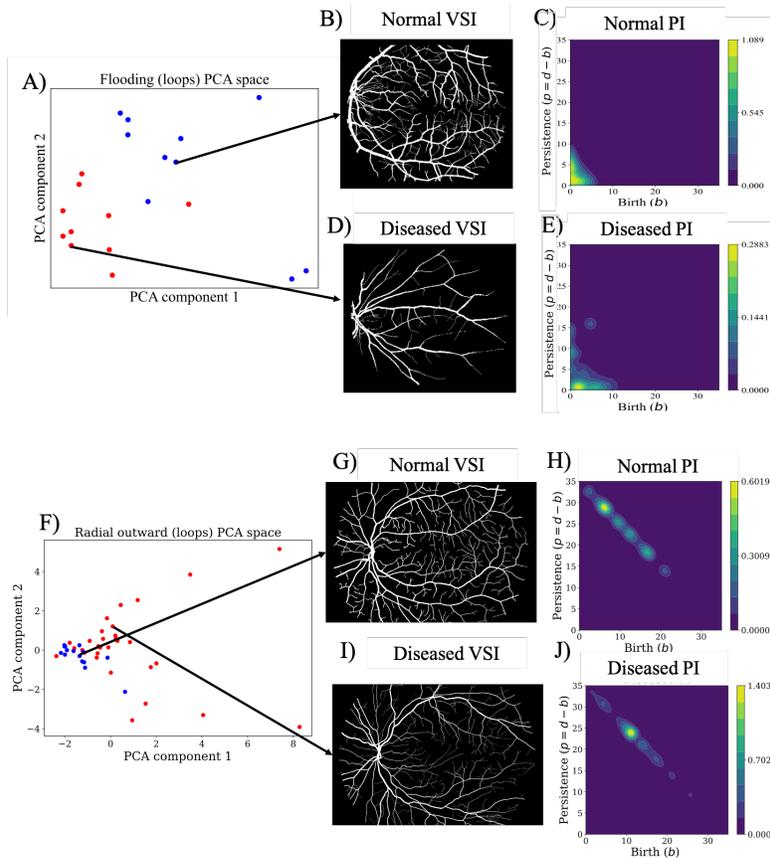

Supplementary Figure 3: Persistence images (PIs) partition the STARE expert 2 and HRF datasets. A) Plotting the dimension-reduced flooding (loops) PIs along the first two principal components for the STARE expert 2 dataset. B) Representative Normal VSI from the STARE expert 2 dataset, and C) Flooding (loops) PI from the representative Normal VSI from the STARE expert 2 dataset. D) Representative Diseased VSI from the STARE expert 2 dataset, and E) Flooding (loops) PI from the representative Diseased VSI from the STARE expert 2 dataset. F) Plotting the dimension-reduced Radial outward (loops) PIs along the first two principal components for the HRF dataset. G) Representative Normal VSI from the HRF dataset, and H) Radial outward (loops)from the representative Normal VSI from the HRF dataset. I) Representative Diseased VSI from the HRF dataset, and J) Radial outward (loops)PI from the representative Diseased VSI from the HRF dataset.